\documentclass[twocolumn]{emulateapj}

\usepackage{txfonts}

\shorttitle{The dependence of star formation on galaxy stellar mass}
\shortauthors{Zheng et al.}

\begin{document}

\title{The Dependence of Star Formation on Galaxy Stellar Mass}


\author{Xian Zhong Zheng,\altaffilmark{1} Eric F. Bell,\altaffilmark{1}
  Casey Papovich,\altaffilmark{2} Christian Wolf,\altaffilmark{3}
  Klaus Meisenheimer,\altaffilmark{1}
  Hans-Walter Rix,\altaffilmark{1} George H. Rieke\altaffilmark{2}
  and Rachel Somerville\altaffilmark{1}}

\altaffiltext{1} {Max-Planck Institut f\"ur Astronomie, K\"onigstuhl 17,
  D-69117 Heidelberg, Germany; xzzheng@pmo.ac.cn.}
\altaffiltext{2} {Steward Observatory, University of Arizona, 933 N Cherry
  Ave, Tucson, AZ 85721.} 
\altaffiltext{3} {Department of Physics, University of Oxford, Keble Road,
  Oxford, OX1 3RH, UK.}

\begin{abstract}

We combine {\it Spitzer} 24\,$\micron$ observations with data from the 
COMBO-17 survey for $\sim$15,000\ $0.2<z\leq 1$ galaxies to determine how 
the average star formation rates (SFR) have evolved for galaxy 
sub-populations of different stellar masses. In the determination 
of $<$SFR$>$ we consider both the ultraviolet (UV) and the infrared (IR) 
luminosities, and account for the contributions of galaxies that are 
individually undetected at 24\,$\micron$ through image stacking. 
For all redshifts we find that higher-mass galaxies have substantially 
lower specific SFR, $<$SFR$>$/$<$M$_\ast$$>$, than lower-mass ones. 
However, we find the striking result that the rate of decline 
in cosmic SFR with redshift is nearly the same for massive and 
low-mass galaxies, i.e. {\it not} a strong function of stellar mass.  
This analysis confirms one version of what has been referred to as  
`downsizing', namely that the epoch of major mass build-up in massive 
galaxies is substantially earlier than the epoch of mass build-up 
in low-mass galaxies. Yet it shows that star formation activity 
is {\it not} becoming increasingly limited to low-mass galaxies 
towards the present epoch.
We argue that this suggests that heating by AGN-powered radio jets 
is not the dominant mechanism responsible for the decline in cosmic SFR 
since $z\sim 1$, which is borne out by comparison with semi-analytic
 models that include this effect.
\end{abstract}


\section{Introduction}\label{intro}

A key observable statistic that describes the evolution of the galaxy population is the average SFR as 
a function of epoch and of galaxy stellar mass.  Over the last decade, 
much progress has been made towards delineating, and in part understanding, the relationship between 
star formation history and stellar mass.  ``Archaeological'' studies of present-day galaxies 
demonstrate a strong correlation between star 
formation history and present-day stellar mass 
in the sense that the bulk of stars now in massive galaxies must have formed at earlier
epochs than stars now in less massive galaxies \citep[e.g.,][]{Kauffmann03,Thomas05,Gallazzi05,Panter06}.  Surveys of cosmologically distant galaxies have revealed a related trend, often referred to as ``downsizing'' in that context.  Downsizing most commonly refers to the observation that intense star formation ($M_\ast/SFR \ll t_{\rm Hubble}$) becomes increasingly limited to systems of lower and lower mass  as redshift decreases \citep{Cowie96,Chapman03,Juneau05,Bauer05,Feulner05,Daddi05,Bell05,Papovich06}.  Recently, this phenomenology has been recast in more quantitative terms 
through study of the specific SFR (i.e., SFR per unit stellar mass) as a 
function of redshift, finding that at $z \la 1$, low-mass galaxies
appear to be forming stars at a higher rate per unit mass than their more massive cousins \citep{Brinchmann04,Bauer05,Noeske07,Feulner05}.

There are two important parts of this empirical picture that require clarification.  Firstly, many studies use SFR
indicators which are highly susceptible to dust extinction (e.g., 
[O\,{\sc ii}] or UV-optical SED-derived SFRs; \citealp{Cowie96,Juneau05,Bauer05,Feulner05}).
Secondly, all studies to date have relied on individual detections to allow SFR derivation. Neglecting SFR contributions from galaxies with individually undetectable star formation tracers biases the census of SFR at all masses. However, the bias is especially severe
at low masses where the SFR limiting sensitivity implies that all low-mass galaxies with 
individual SFR detections have extremely high specific SFR.

In this letter, we present an analysis of Spitzer 24\,$\micron$ observations of 
two COMBO-17 fields designed to circumvent these two 
key limitations.  We estimate SFR by combining indicators of obscured and unobscured 
star formation, and attempt to derive average SFRs through stacking of galaxy sub-samples with individually-undetected SFR indicators.  
In \S 2, we summarize the data and galaxy samples. 
In \S 3, we describe our SFR measurements. 
In \S 4, we present the contribution of different bins in stellar mass
to the cosmic SFR, and present the evolution of specific SFR 
with lookback time, again as a function of stellar mass.
We discuss our results in \S 5.  
We adopt a  cosmology with H$_0$\,=\,70\,km\,s$^{-1}$\,Mpc$^{-1}$,
$\Omega_{\rm M}$\,=\,0.3 and $\Omega_{\Lambda}$\,=\,0.7.

\section{The Data and Galaxy Samples}

The COMBO-17 survey \citep{Wolf03} has imaged three $30\arcmin \times 30\arcmin$ fields in
five broad and 12 intermediate optical bands from 350 to 930\,nm with 
the goal of deriving high-quality photometric redshifts\footnote{Redshift uncertainties are $\delta z/(1+z)\sim 0.02$ at the median galaxy magnitude of $m_{\rm R} \sim 22$ \citep{Wolf04}.}
for $>10\,000$ galaxies with $m_{\rm  R}<24$\,mag in each
field.  The rest-frame luminosities in the
$U$, $B$ and $V$ band, and a synthetic UV2800 band centered at 280\,nm have been derived for all COMBO-17 galaxies with photo-z's.  By fitting the COMBO-17 multi-band photometry with a library of SED templates, 
\citet{Borch06} estimated the galaxies' stellar mass-to-light ratio ($M/L$), and hence stellar mass  for all $m_{\rm  R}<24$ galaxies based on a \citet{Kroupa93} initial mass function (IMF; such an IMF has 
stellar masses similar to within 10\% compared to those
derived using a \citealt{Chabrier03} or \citealt{Kroupa01} IMF). From these data, \citet{Borch06} constructed stellar mass
functions in four even redshift slices between $z=0.2$ and $z=1$. 

Two of three COMBO-17 fields, the extended Chandra Deep field South (CDFS)
and the Abell 901/902 (A901) have been observed by the Multiband Imaging Photometer 
on {\it Spitzer} (MIPS: \citealp{Rieke04}) at 24\,$\micron$. 
 The reduced 24\,$\micron$ images cover an area of $1\degr \times 0\fdg 5$ around
the CDFS \citep{Papovich04} and of $30\arcmin \times 55\arcmin$ in the A901 field \citep{Bell07}. Both images
have a pixel scale of $1\farcs25$\,pixel$^{-1}$, a point spread function (PSF)
with full width at half maximum (FWHM) of $\simeq$\,6$\arcsec$. The IRAF/DAOPHOT package and empirically-determined PSFs are used to simultaneously fit multiple sources to the 24\,$\micron$ images and derive total flux for each detected object to a depth of 83\,$\mu$Jy at the 5\,$\sigma$ level (to a completeness of 80\%; see also \citealp{Papovich04} for details of data reduction, source detection
and photometry). To match these 24\,$\micron$ catalogs to the COMBO-17 galaxy samples, a position tolerance of $2\farcs2$ is adopted. If more than one galaxy exist within the tolerance, the nearest one is selected. The overlap area between COMBO-17 and MIPS imaging contains 9785 and 12995 galaxies at $m_{\rm R} < 24$ in the CDFS ($\sim$800 square arcminutes) and the A901 field ($\sim$850 square arcminutes), respectively. Of them, 1735 (1872) are individually detected at 24\,$\micron$. Only $<$4\% (6\%) of the 24\,$\micron$-detected sources, namely 64 (108), have multiple (mostly two) COMBO-17 galaxies within the tolerance. 
We select galaxies with stellar mass $M_\ast > 10^9$\,M$_\odot$ in the redshift range $0.2<z\leq 1$, resulting in a sample of 6893 (8184) galaxies in the CDFS (A901) field. Of the 6893 (8184) galaxies, 1201 (1250) are individually detected at  24\,$\micron$. The {\it average} 24\,$\micron$ flux of the galaxies individually undetected by MIPS will be estimated by stacking, as described in the next section.
The samples are complete at mass limit $\sim$1, 1.8, 3, 
6$\times 10^{10}$\,M$_\odot$ for redshift $z\sim$0.3, 0.5, 0.7, 0.9 \citep{Borch06}.

\begin{figure}
\centering
\includegraphics[width=0.48\textwidth]{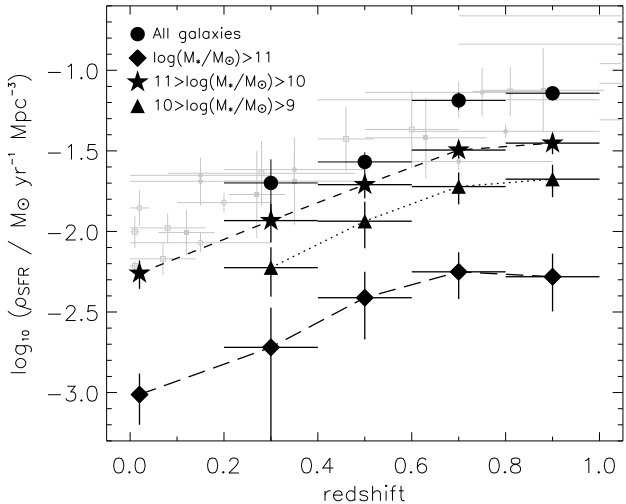}
\caption{The volume-averaged SFR as functions of redshift, for galaxy samples of differing mean stellar masses. The average SFR is derived from the average bolometric
  luminosity (UV+IR) for galaxy subsets limited in mass and redshift. 
  The gray symbols are existing measurements
  of the overall SFR density in the literature, adapted from \citet{Hopkins04}.
  The solid circles show the overall SFR density
  derived from the SFR functions given in \citet{Bell07}. 
  The data imply that the fraction of the total SFR contributed by galaxy sub-samples of different stellar masses has not changed since $z\sim 1$, despite the factor $\sim$7 decline in the total SFR in that interval.
}
\label{sfrds}
\end{figure}

Owing to mismatches in depth between wide-area IR surveys (i.e., IRAS) and well-characterized 
optical galaxy surveys (i.e., the SDSS and 2dFGRS), it is challenging to produce
an appropriate and well-defined local comparison sample.  We have attempted to produce an adequate
local control sample by collecting a sample of 2177 galaxies from 
the NASA/IPAC Extragalactic Database (NED) with 2MASS $K$-band magnitude 
$K<12$ in the volume of $1500 \leq cz /({\rm km\,s^{-1}}) \leq 3000$ and 
Galactic latitude $b> 30\degr$. 
Assuming a $K$-band $M/L$ of 0.6\,M$_\odot$/L$_\odot$ and a 
Kroupa IMF, stellar masses were derived from the 
$K$-band absolute magnitude. The SFR is estimated from the total IR luminosity,
as determined from IRAS observations.  The local sample is not a complete volume-limited
sample. However, we estimate that above $M_\ast > 10^{10}$\,M$_\odot$ the sample
is $\sim$65\% complete with an incompleteness which is largely geometric in origin (because
some areas of the sky were not spectroscopically surveyed as thoroughly as others).  
If indeed the incompleteness is primarily geometric in origin, the average
SFRs at a given stellar mass should not be biased.\footnote{At any rate, inspection 
of Fig.\ \ref{sfrds} and Fig.\ \ref{ssfr} shows that none of our conclusions actually
depend on the $z \sim 0$ data; the purpose of its inclusion is simply to extend the 
trends to $z \sim 0$.} 
See \citet{Bell05} for more details about the local sample selection and completeness.

\begin{deluxetable}{cccc}
  \centering \tabletypesize{\scriptsize} \tablewidth{0pt}
  \tablecaption{The cosmic star formation history split by galaxy stellar mass. 
  \label{table}}
\tablehead{ & \multicolumn{3}{c}{$\log (\rho_{\rm SFR}$/M$_\odot$\,yr$^{-1}$\,Mpc$^{-3}$)}  \\
            \cline{2-4} \\
        $z$ &  $10^{9}<\frac{M_\ast}{{\rm M}_\odot} \leq 10^{10}$ & $10^{10}<\frac{M_\ast}{{\rm M}_\odot} \leq 10^{11}$ & $\frac{M_\ast}{{\rm M}_\odot} >10^{11}$ }

\startdata
$\sim$0.01 & ...                     & $-2.26^{+0.08}_{-0.10}$ & $-3.01^{+0.13}_{-0.19}$ \\
0.2 -- 0.4   & $-2.22^{+0.13}_{-0.18}$ & $-1.93^{+0.10}_{-0.14}$ & $-2.72^{+0.25}_{-0.63}$ \\
0.4 -- 0.6   & $-1.94^{+0.12}_{-0.17}$ & $-1.71^{+0.07}_{-0.08}$ & $-2.41^{+0.16}_{-0.26}$ \\
0.5 -- 0.8   & $-1.72^{+0.09}_{-0.11}$ & $-1.49^{+0.05}_{-0.05}$ & $-2.25^{+0.12}_{-0.17}$ \\
0.8 -- 1.0   & $-1.68^{+0.09}_{-0.11}$ & $-1.45^{+0.06}_{-0.07}$ & $-2.28^{+0.14}_{-0.22}$ \\
\enddata 
\end{deluxetable}

\section{SFR Determination}

We  split the sample galaxies 
into four redshift slices evenly covering the range from $z=0.2$ to $z=1$. 
In each redshift slice, sample galaxies are divided into five stellar mass bins: 
four bins with a bin width of 0.5\,dex spanning from $M_\ast = 10^9$\,M$_\odot$ 
to $10^{11}$\,M$_\odot$, and one bin for $M_\ast>10^{11}$\,M$_\odot$.  
Each sub-sample of galaxies defined by optical data 
contains individually-detected and 
individually-undetected objects at 24\,$\micron$. 
Following \citet{Zheng06} we estimated the total
24\,$\micron$ flux for the sub-sample of individually-undetected objects through 
stacking. The total 24\,$\micron$ flux of the sub-sample 
is then obtained by adding the individual fluxes from the detected objects 
and stacked flux from the individually-undetected objects. 
Our SFR indicator, the total IR luminosity (8-1000\,$\micron$), is extrapolated 
from the 24\,$\micron$ luminosity using three sets of local IR 
SED templates from \citet{Chary01}, 
\citet{Dale02} and \citet{Lagache04}. 
In \citet{Zheng07}, we use stacking of IR-bright galaxies
with $0.6 < z < 0.8$ at 70\,$\micron$ and 160\,$\micron$ to show that the average IR
SED of an IR-bright galaxy is spanned by these local template sets, lending 
credibility to the total IR luminosity estimated here to within a factor of two.

We use the bolometric luminosity produced by young stars (UV+IR) to estimate 
the SFR, a procedure that accounts for both obscured and unobscured star formation.  
The total UV luminosity (1216\,\AA\--3000\,\AA) is estimated 
from the COMBO-17 2800{\AA} rest-frame luminosity $L_{\nu,2800}$ as 
$L_{\rm UV}$\,=\,1.5\,$\nu L_{\nu,2800}$. 
We used the conversion from \citet{Bell05} to convert the UV+IR
into SFR.  The SFR is calibrated to a \citet{Chabrier03} IMF.
Using the sum of stellar masses for each sub-sample from \citet{Borch06}, we calculated the 
average specific SFR, i.e. $<SFR>/<M_\ast>$, for 
all redshift and stellar mass bins.
Formal uncertainties were estimated from bootstrapping. 
For each redshift and stellar mass bin, the average 
between the CDFS and A901 was taken as the final result; 
the difference between the two is taken as a measure of field-to-field 
variance and included into the uncertainties.  
The estimates of both SFRs and stellar masses
are affected by systematic errors, that other studies have estimated to be  
$<0.3$\,dex.  These systematic uncertainties are not included
in our error bars, but are substantially smaller than the dynamic range
of the trends discussed in this Letter.
Due to the COMBO-17 $R$-band selection ($m_{\rm  R}<24$), our sample galaxies
are incomplete for low-mass bins at $z>0.4$, where low-mass red galaxies 
with little blue radiation from newly-formed stars
fail to make the $R$-band selection cut but low-mass blue galaxies containing more on-going star formation remain available for the selection.  
The selection effect may lead to a potential overestimate of the
specific SFR for these incomplete bins.  The specific SFR at $z\sim 0$ was estimated 
using the local sample only for
three high-mass bins where the sample is representative.  

\section{$<SFR>\,(z)$ as a Function of Galaxy Stellar Mass}

In the previous section, we have determined the average SFRs for galaxies
split into stellar mass and redshift bins.  In order to determine the
contribution of different stellar mass bins to the cosmic (average) SFR,
we must combine the average SFRs at a given mass with the space density of
galaxies in that same mass range, as a function of redshift.  For this
purpose, we adopt the stellar mass functions of \citet{Borch06} for $0.2 <
z \le 1$, and \citet{Bell03} for $z \sim 0$, although the results are
unaffected to within the uncertainties if we adopt stellar mass functions
from, e.g., \citet{Drory04}, \citet{Fontana06} or \citet{Bundy06} instead.
We calculate the volume-averaged SFR over the redshift
range $0<z\leq 1$ for three galaxy stellar mass bins: 
$M_\ast/{\rm M}_\odot >10^{11}$, $10^{10} - 10^{11}$ and $10^9 - 10^{10}$.
The stellar mass functions from \citet{Borch06} are derived
from $\sim$25\,000 COMBO-17 galaxies selected from three separated $30\arcmin
\times 30\arcmin$ fields. Therefore, the field-to-field variance is somewhat 
suppressed in this way (as one can see from comparison of the total SFR, derived from 
two COMBO-17 fields by \citet{Bell07} compared to the smooth trends
followed by our mass-binned cosmic SFRs).
The local data point is not available for the stellar mass bin 
$10^9< M_\ast/{\rm M}_\odot \leq 10^{10}$.

The results are listed in Table~\ref{table} and shown in Fig.~\ref{sfrds}, along with
current measurements of the cosmic SFR density available in the literature
\citep{Hopkins04}. \citet{Bell07}
constructed SFR functions using the same data sets and SFR estimator 
(UV+IR) as adopted here. The data points given by their SFR functions 
are shown also in this plot.  
Our estimate of the cosmic SFR density, i.e. the sum of the volume-averaged 
SFRs from the three stellar mass bins (contribution from galaxies of 
stellar mass $< 10^{9}$\,M$_\odot$ is negligible), is in good agreement
with Bell et al.'s and other measurements from the literature. 

Fig.~\ref{sfrds} shows that the volume-averaged SFR of galaxies in 
each of the three stellar mass bins decreases towards the present day 
with an indistinguishably similar slope since $z=1$. 
In other words, the contributions to the cosmic SFR density from galaxies 
of different stellar mass ranges has not changed with cosmic epoch during 
the last 8\,Gyr.  Galaxies in the 
$10^{10}< M_\ast/{\rm M}_\odot \leq 10^{11}$ contain the bulk of 
the overall star formation at all cosmic epochs which we have examined.
These galaxies also dominate the cosmic stellar mass density 
\citep{Borch06}. A smaller part of the overall 
star formation takes place in galaxies with
$10^{9}< M_\ast/{\rm M}_\odot \leq 10^{10}$.  Relatively little star formation is associated 
with massive galaxies with $M_\ast > 10^{11}$\,M$_\odot$.

\begin{figure}
\centering
\includegraphics[width=0.48\textwidth]{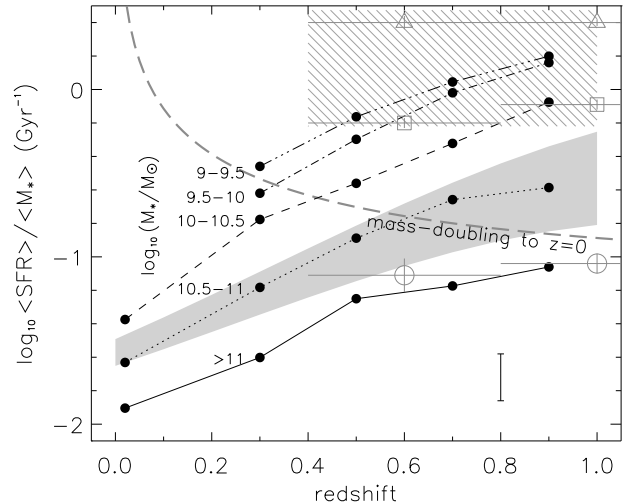}
\caption{The specific SFR (SFR per unit stellar mass), $<SFR>/<M_\ast>$, as function of redshift for galaxy samples of differing stellar masses.  The average SFR is derived from the average bolometric
  luminosity (UV+IR) for galaxy subsets limited in mass and redshift. 
  The data points in the hatched region are incomplete due to the COMBO-17 source selection   ($m_{\rm  R}<24$). 
  The gray strip illuminates the overall specific SFR, i.e. the ratio of
  the cosmic SFR density to the cosmic stellar mass density, following \citet{Hopkins04} and \citet{Borch06}. 
  Local data points are estimated for three high-mass bins using a local
  galaxy sample with near-IR and IRAS observations. The errorbar represents 
  the typical uncertainty of our specific SFR determinations. 
  The dark-gray 
  dashed line is the specific SFR required to double the mass of a galaxy by 
  the present day. Data points above the mass-doubling line represent galaxies
  in intense star formation mode and those below the line represent galaxies 
  in quiescent star formation mode. For all galaxy sub-samples of different 
  stellar masses, the specific SFR increases with redshift out to $z\sim 1$. 
  At all redshifts, more massive galaxies always have lower specific SFR than 
  less massive galaxies. 
These relations delineate a picture in which lower-mass galaxies transition from an intense into a quiescent star 
formation phase since $z \sim 1$, whereas most massive galaxies have 
formed stars quiescently across this epoch.
Open symbols show results from Feulner et al. (\citeyear{Feulner05}) based on SFR indicator rest-frame UV luminosity for stellar mass bins (converted to a Kroupa IMF) $10^{8.25}< M_\ast/{\rm M}_\odot < 10^{9.25}$ ({\it triangles}), $10^{9.25}< M_\ast/{\rm M}_\odot < 10^{10.25}$ ({\it squares}) and $10^{10.25}< M_\ast/{\rm M}_\odot < 10^{11.25}$ ({\it circles}).
}
\label{ssfr}
\end{figure}

Fig.~\ref{ssfr} presents our measurements of specific SFR, i.e. SFR per unit stellar mass. 
Data points of the same stellar mass bin are linked together, 
showing the redshift evolution of the specific SFR. Data points of the 
same redshift delineate the specific SFR as a function of stellar mass. 
Those enclosed in the shaded region are in bins affected by serious incompleteness due to the COMBO-17 flux limit. 
For comparison, we overplotted the average specific SFR for all galaxies, 
i.e. the ratio of the cosmic SFR density to the cosmic stellar mass density, 
calculated using \citet{Hopkins04} and \citet{Borch06}. 
As shown in Fig.~\ref{ssfr}, the average specific SFR for a
galaxy stellar mass bin increases rapidly with redshift, in qualitative
agreement with previous studies \citep[e.g.,][]{Bauer05,Feulner05}. Compared to our results, SFR estimates from rest-frame UV luminosity \citep[open symbols in Fig.~\ref{ssfr};][]{Feulner05} produce lower specific SFRs for high-mass galaxies and higher specific SFRs for low-mass galaxies due to the reasons mentioned in \S~\ref{intro}.

Yet, the change in specific SFR is nearly equally strong for all galaxy mass bins. At all redshifts, the population-averaged specific SFR of less massive galaxies is higher than that of more massive galaxies. Note that, as expected, galaxies of stellar mass around  $10^{10.5} < M_\ast/{\rm M}_\odot \leq 10^{11}$
have specific SFR comparable to the `cosmic average' specific SFR, 
as they contribute most to both the cosmic average SFR and stellar mass budget. 

\section{Discussion and Summary}

We combine {\it Spitzer} 24\,$\micron$ observations with 
COMBO-17 data to explore the evidence for the `downsizing' of the star formation activity since $z\sim 1$. Specifically, we estimate SFR from the UV (for unobscured star formation) 
and the IR luminosities (for obscured star formation) for a sample 
of $\sim$15\,000 galaxies selected from the CDFS and A901 fields. 
Through image stacking, we account for the SFR contributions of 
otherwise undetectable galaxies.
Our measurements of average SFR for galaxy subsets limited in mass and 
redshift account for contributions from both star-forming galaxies and 
quiescent galaxies 
in each subset. We find that the relative drop in SFR from $z \sim 1$ 
to the present day is nearly identical for different sub-samples that 
span a range of $\sim$100 in stellar mass. That is,
all galaxy samples, independent of mass, 
exhibit the same gradual decline of star formation 
(and specific star-formation) over the last 8 Gyr.  
At all redshifts, we find that high-mass galaxies
have substantially lower specific SFR than their low-mass cousins.

While many of the qualitative features of this paper have been 
understood for some time \citep[e.g.,][]{Brinchmann00,Bauer05,Juneau05}, 
this Letter quantifies the relationships between specific SFR, galaxy stellar mass and redshift.  `Archaeological' downsizing --- that the stars in massive
galaxies formed at earlier times than the stars in low-mass galaxies ---
is clearly recovered in this paper, in Fig.\ \ref{ssfr}.  On average, low-mass galaxies have 
specific SFRs capable of more than doubling their stellar mass over the Hubble time (at
all redshifts), whereas galaxies with $M_\ast > 10^{11}$\,M$_{\odot}$ are not capable of such 
growth at $z\la 1$ from {\it in situ} star formation alone.  

Arguably the most striking and new contribution of this Letter is that
the decline in SFR from $z \sim 1$ to the present occurs at the same rate in
galaxies of all masses.  A number of mechanisms have been 
postulated to drive the demise of star formation since $z \sim 1$: 
AGN feedback, the declining galaxy merger rate, gas consumption, 
changes in the availability of cooling or infalling fresh gas, and 
a host of other possibilities.  
We speculate that the declining supply of new gas falling in from 
the cosmic web at $z \la 1$ seems an attractive candidate for 
driving much of the phenomenology that we observe \citep{Noeske07}.
The roughly constant rate of decline of SFR and SSFR for galaxies of 
widely differing masses is in contrast to the predictions of theoretical 
models in which cooling and hence star formation is quenched by heating 
due to radio-loud AGN (\citealp[e.g.,][]{Croton06,Bower06}; but see also \citealt{Neistein06}). This process 
is more important in more massive galaxies for two reasons. First, more 
massive galaxies are more likely to host massive black holes, and more 
massive black holes are generally assumed to be capable of producing 
more energy and stronger heating. Second, it is generally assumed in 
such models that AGN can only heat gas that is cooling from a 
quasi-hydrostatic hot halo, which form in large mass halos 
($M_{\rm halo} >\, \sim$10$^{12}$\,M$_\odot$) as opposed to the ``cold mode'' 
flows that are dominant in lower 
mass halos. For example, the models of \citet{Bower06} predict a 
decline of a factor of $\sim$8 from $z\sim 1$ to the present, in the 
SFR contributed by galaxies more massive than $10^{10.8}$\,M$_\odot$ (see 
their Figure 8), while the SFR contributed by less 
massive galaxies declines by a factor of about 2--4 over this same 
interval. We find, using an independent model with 'radio-mode' heating 
by AGN (Somerville et al. in preparation), that the specific star formation 
rate in galaxies with stellar mass greater than $10^{11}$\,M$_\odot$ has 
declined by a factor of $\sim$15 since $z\sim 1$, while the specific SFR 
in the lower mass 
bins (the same bins used here), declines by only a factor of 3--5. The 
precise rate of decline of the SFR as a function of 
galaxy mass is no doubt sensitive to the details of the physics of star 
formation, but the prediction of a strongly differential  `quenching' of 
star formation by AGN feedback in the form of radio heating, as it is 
currently implemented, seems inevitable. Our results therefore suggest 
that radio mode heating by AGN is not the dominant mechanism responsible 
for the decrease in cosmic SFR from $z\sim 1$ to the present.


\acknowledgments

E.\ F.\ B.\ was supported by the Emmy Noether Programme of the Deutsche Forschungsgemeinschaft.
C.\ W.\ was supported by a PPARC Advanced Fellowship. We thank the anonymous referee for helpful suggestions that improved our manuscript. 
This research has made use of the NASA/IPAC Extragalactic Database (NED) which is operated by the Jet Propulsion Laboratory, California Institute of Technology, under contract with the National Aeronautics and Space Administration.


\begin{thebibliography}{}

\bibitem[Bauer et al.(2005)]{Bauer05} Bauer, A. E., Drory, N., Hill, G. J.,
  \& Feulner, G. 2005, ApJ, 621, L89
\bibitem[Bell et al.(2003)]{Bell03} Bell, E. F., McIntosh, D., Katz, N., \&
  Weinberg, M. D.  2003, ApJS, 149, 289
\bibitem[Bell et al.(2005)]{Bell05} Bell, E. F., et al. 2005, ApJ, 625, 23
\bibitem[Bell et al.(2007)]{Bell07} Bell, E. F., Zheng, X. Z., Papovich, C., Borch, A., Wolf, C., \& Meisenheimer, K. 2007, ApJ, submitted
\bibitem[Borch et al.(2006)]{Borch06} Borch, A., et al. 2006, A\&A, 453, 869
\bibitem[Bower et al.(2006)]{Bower06} Bower, R. G., et al. 2006, MNRAS, 370, 645
\bibitem[Brinchmann \& Ellis(2000)]{Brinchmann00} Brinchmann, J., \& Ellis,
  R. S. 2000, ApJ, 536, L77
\bibitem[Brinchmann et al.(2004)]{Brinchmann04} Brinchmann, J., et al. 2004,
  MNRAS, 351, 1151 
\bibitem[Bundy et al.(2006)]{Bundy06} Bundy, K., et al. 2006, ApJ, 651, 120 
\bibitem[Chabrier(2003)]{Chabrier03} Chabrier, G. 2003, ApJ, 586, L133
\bibitem[Chapman et al.(2003)]{Chapman03} Chapman, S. C., Blain, A. W.,
  Ivison, R. J., \& Smail, I. R. 2003, Nature, 422, 695 %
\bibitem[Chary \& Elbaz(2001)]{Chary01} Chary, R., \& Elbaz, D. 2001, ApJ,
  556, 562
\bibitem[Cowie et al.(1996)]{Cowie96} Cowie, L. L., Songaila, A., \& Hu, E. M. 1996, AJ, 112, 839
\bibitem[Croton et al.(2006)]{Croton06} Croton, D. J., et al. 2006, MNRAS, 365, 11  
\bibitem[Daddi et al.(2005)]{Daddi05} Daddi, E., et al. 2005, ApJ, 631, L13
\bibitem[Dale \& Helou(2002)]{Dale02} Dale, D. A., \& Helou, G. 2002, ApJ, 576, 159
\bibitem[Drory et al.(2004)]{Drory04} Drory, N., et al. 2004, ApJ, 608, 742
\bibitem[Feulner et al.(2005)]{Feulner05} Feulner, G., Gabasch, A., Salvato, M., Drory, N., Hopp, U., \& Bender, R. 2005, ApJ, 633, L9
\bibitem[Fontana et al.(2006)]{Fontana06} Fontanta, A., et al. 2006,  A\&A, 2006, 459, 745
\bibitem[Gallazzi et al.(2005)]{Gallazzi05} Gallazzi, A., Charlot, S., Brinchmann, J. White, S. D. M., \& Tremonti, C. A. 2005, MNRAS, 362, 41 
\bibitem[Hopkins(2004)]{Hopkins04} Hopkins, A. M., 2004, ApJ, 615, 219
\bibitem[Lagache et al.(2004)]{Lagache04} Lagache, G., et al. 2004, ApJS,
  154, 112
\bibitem[Kauffmann et al.(2003)]{Kauffmann03} Kauffmann, G., et al. 2003, MNRAS, 341, 54 
\bibitem[Kroupa(2001)]{Kroupa01} Kroupa, P. 2001, MNRAS, 322, 231 %
\bibitem[Kroupa et al.(1993)]{Kroupa93} Kroupa, P., Tout, C. A., \& Gilmore, G. 1993, MNRAS, 262, 545
\bibitem[Juneau et al.(2005)]{Juneau05} Juneau, S., et al. 2005, ApJ, 619, L135
\bibitem[Neistein et al.(2006)]{Neistein06} Neistein, E., van den Bosch, F. C., \& Dekel, A. 2006, MNRAS, 372, 933
\bibitem[Noeske et al.(2007)]{Noeske07} Noeske, K. G., et al. 2007, ApJ, in press (astro-ph/0701924)
\bibitem[Panter et al.(2006)]{Panter06} Panter, B., Jimenez, R., Heavens,
  A. F., \& Charlot, S. astro-ph/0608531
\bibitem[Papovich et al.(2004)]{Papovich04} Papovich, C., et al. 2004, ApJS,
  154, 70 
\bibitem[Papovich et al.(2006)]{Papovich06} Papovich, C., et al. 2006, ApJ,
  640, 92
\bibitem[Rieke et al.(2004)]{Rieke04} Rieke, G. H., et al. 2004, ApJS, 154, 25
\bibitem[Thomas et al.(2005)]{Thomas05} Thomas, D., Maraston, C., Bender, R.,
  \& de Oliveira, C. M. 2005, ApJ, 621, 673 
\bibitem[Wolf et al.(2003)]{Wolf03} Wolf, C., et al. 2003, A\&A, 401, 73
\bibitem[Wolf et al.(2004)]{Wolf04} Wolf, C., et al.  2004, A\&A, 421, 913     
\bibitem[Zheng et al.(2006)]{Zheng06} Zheng, X. Z., Bell, E. F., Rix, H.-W., Papovich, C., Le Floc'h, E., Rieke, G. H., \& P\'erez-Gonz\'alez, P. G.  2006, ApJ, 640, 784
\bibitem[Zheng et al.(2007)]{Zheng07} Zheng, X. Z., Dole, H., Bell, E. F., Le Floc'h, E., Rieke, G., H., Rix, H.-W., \& Schiminovich, D.  2007, ApJ, submitted
\end{thebibliography}
\end{document}